\documentclass[floats,floatfix,showpacs,amssymb,prd,superscriptaddress,twocolumn,aps]{revtex4}
\usepackage{graphicx, epsfig, amssymb} 
\usepackage{amsmath, amsfonts}
\usepackage{gensymb}
\usepackage{bm} 

\usepackage[linktocpage]{hyperref}
\usepackage[caption=false]{subfig}
\usepackage[usenames]{color}

\def\be{\begin{equation}}
\def\ee{\end{equation}}
\def\beq{\begin{eqnarray}}
\def\eeq{\end{eqnarray}}

\newcommand{\bea}{\begin{eqnarray}}
\newcommand{\eea}{\end{eqnarray}}
\newcommand{\ben}{\begin{enumerate}}
\newcommand{\een}{\end{enumerate}}
\newcommand{\bi}{\begin{itemize}}
\newcommand{\ei}{\end{itemize}}

\pdfoutput=1

\begin{document}

\title{Fundamental photon orbits: \\  black hole shadows and spacetime instabilities}

 \author{Pedro V. P. Cunha}
 \affiliation{
   Departamento de F\'\i sica da Universidade de Aveiro and CIDMA, 
   Campus de Santiago, 3810-183 Aveiro, Portugal.
 }
   \affiliation{
   CENTRA, Departamento de F\'\i sica, Instituto Superior T\'ecnico \\  Universidade de Lisboa, Avenida Rovisco Pais 1, 1049, Lisboa, Portugal.
 }

 \author{Carlos~A.~R.~Herdeiro}
   \affiliation{
   Departamento de F\'\i sica da Universidade de Aveiro and CIDMA, 
   Campus de Santiago, 3810-183 Aveiro, Portugal.
 }

 \author{Eugen Radu}
   \affiliation{
   Departamento de F\'\i sica da Universidade de Aveiro and CIDMA, 
   Campus de Santiago, 3810-183 Aveiro, Portugal.
 }


\date{\today}

\begin{abstract}
The standard  Black Holes (BHs) in General Relativity, as well as other ultra-compact objects (with or without an event horizon) admit planar circular photon orbits. These \textit{light rings} (LRs) determine several spacetime properties. For instance, stable LRs trigger instabilities and, in spherical symmetry, (unstable) LRs completely determine BH shadows.  In generic stationary, axi-symmetric spacetimes, \textit{non-planar} bound photon orbits may also exist, regardless of the integrability properties of the photon motion. We suggest a classification of these  \textit{fundamental photon orbits} (FPOs) and, using Poincar\'e maps, determine a criterion for their stability. For the Kerr BH, all FPOs are unstable (similarly to its LRs) and completely determine the Kerr shadow.  But in non-Kerr spacetimes, stable FPOs may also exist, even when all LRs are unstable, triggering new instabilities. We illustrate this for the case of Kerr BHs with Proca hair, wherein, moreover, qualitatively novel shadows with a cuspy edge exist, a feature that can be understood from the interplay between stable and unstable FPOs. FPOs are the natural generalisation of LRs beyond spherical symmetry and should generalise the LRs key role in different spacetime properties. 
\end{abstract}




\maketitle
\noindent{\bf {\em Introduction.}} \textit{Light rings} (LRs), $i.e.$ circular photon orbits, are an extreme form of light bending by ultracompact objects (UCOs). They have distinct phenomenological signatures in both the electromagnetic and gravitational wave channels. In the former, LRs are closely connected to the \textit{shadow} of a black hole (BH)~\cite{1973blho.conf..215B,Falcke:1999pj}. This is the absorption cross section of light at high frequencies, an observable that is being targeted by the Event Horizon Telescope~\cite{2006MNRAS.367..905B,2009astro2010S..68D}. In the gravitational wave channel, LRs determine a perturbed BH's early-time ringdown~\cite{1972ApJ...172L..95G}, corresponding to the post-merger part of the recently detected gravitational wave transients by aLIGO~\cite{Abbott:2016blz,Abbott:2016nmj}.  The frequency and damping time of this early-time ringdown are set by the orbital frequency and instability time scale (Lyapunov exponent) of an (unstable) LR.

LRs also define other dynamical properties of UCOs. For horizonless UCOs, LRs often come in pairs, one being stable and the other unstable. The existence of a stable LR has been claimed to imply a spacetime instability~\cite{Keir:2014oka,Cardoso:2014sna}. Finally, LRs impact on our Newtonian intuition for test particle motion: crossing (inwards) a LR swaps the perception of inwards/outwards, and reverses the centrifugal effect of angular motion~\cite{1993SciAm.268R..26A}. 

For spherical UCOs, LRs (which are always planar) are the only bound photon orbits. But for an axisymmetric (and stationary) spacetime more general photon orbits are possible, that neither escape to infinity, nor fall into a BH (if the UCO is a BH). In this letter, we analyse implications, and propose a classification, of this natural generalization of LRs, dubbed \textit{fundamental photon orbits} (FPOs). In particular we argue they can trigger new spacetime instabilities and show they are paramount in understanding the detailed structure of BH shadows.

\begin{widetext}

\begin{figure}[b!]
\begin{center}
\includegraphics[width=0.385\textwidth]{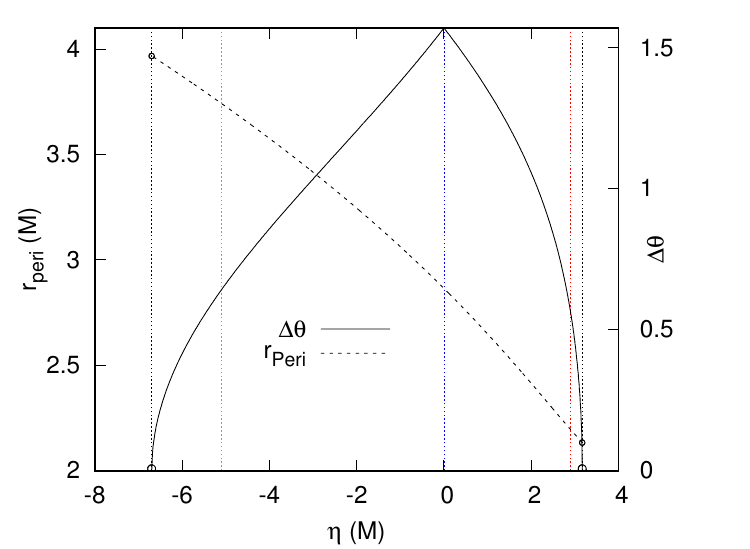}
\includegraphics[width=0.34\textwidth]{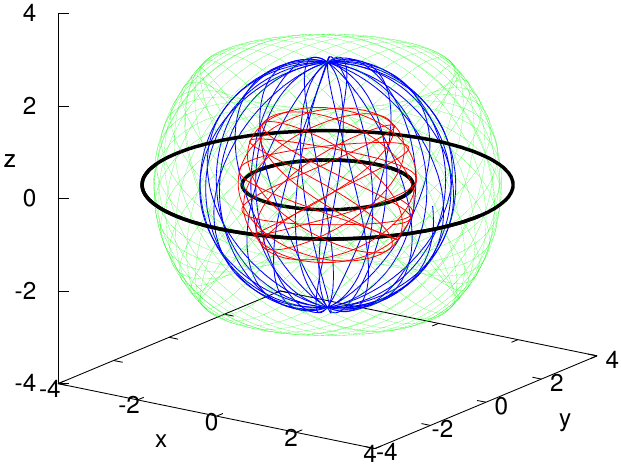}
\includegraphics[width=0.26\textwidth]{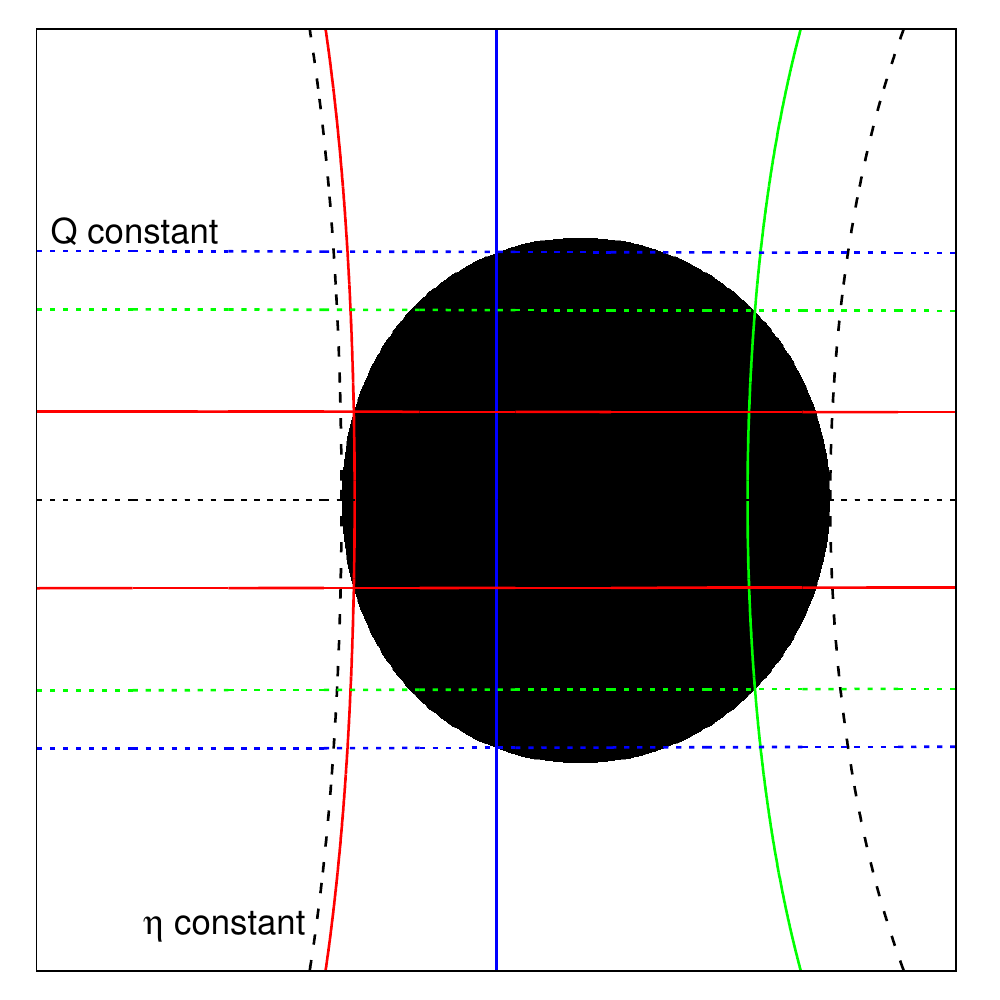}
\end{center}
\caption{\small Kerr(-like) FPOs and shadow, illustrated for a Kerr BH with dimensionless spin $j \simeq 0.820$, $\eta_-=-6.70$, $\eta_+=3.17$. (Left panel) $r_{\rm Peri}$ and $\Delta \theta$ for FPOs $vs.$ $\eta$.  Lines with $\eta={\rm constant}$ take the values of the LRs or 3 selected FPOs, $\eta=-5.10,0,2.90$. (Middle panel) Spatial trajectories of these 3 FPOs and  2 LRs, in Cartesian coordinates defined from BL coordinates.  (Right panel) BH shadow, in the same observations conditions as Fig ~\ref{shadowsnk}. Almost vertical (solid) lines have $\eta={\rm constant}$ and horizontal (dotted) lines have fixed Carter's constant $Q$, both with the values of the 3 selected FPOs. Observe how the FPOs ($\eta,Q$) values correspond to points at the edge of the shadow.  The same colours are used in all panels for the same FPOs.}
\label{kerrfig}
\end{figure}

\end{widetext}

\noindent{\bf {\em FPOs.}} In vacuum General Relativity (GR), the only regular (on and outside an event horizon) UCO is the Kerr solution~\cite{Kerr:1963ud}, wherein geodesic motion is Liouville integrable and separates in Boyer-Lindquist  (BL) coordinates $(t,r_{BL},\theta, \varphi)$~\cite{Carter:1968rr}. In this chart,  FPOs with constant $r_{BL}$ and motion in $\theta$ exist, known as \textit{spherical orbits}~\cite{Teo}. The subset restricted to the equatorial plane are the two LRs, one for co-rotating and one for counter-rotating photons (with respect to the BH), both converging at $r_{BL}=3M$ in the Schwarzschild BH (mass $M$) limit~\cite{Bardeen:1972fi}. 
Spherical orbits are related to the ringdown modes in BH perturbation theory~\cite{cardoso2009geodesic} and completely determine the Kerr BH shadow ($cf.$ Fig.~\ref{kerrfig}). These are the most general FPOs in Kerr~\footnote{This follows from the inexistence of more than one radial turning point for photon motion on the Kerr solution.}, all of them unstable.

For generic stationary and axisymmetric spacetimes, we define FPOs as follows:\\
\textit{Definition:} let $s(\lambda):\mathbb{R}\to\mathcal{M}$ be an affinely parameterised null geodesic, mapping  the real line to the space-time manifold $\mathcal{M}$. $s(\lambda)$ is a FPO if it is restricted to a compact spatial region -- it is a bound state -- and if there is a value $T>0$ for which $s(\lambda)=s(\lambda+T), \forall\,\lambda\in\mathbb{R}$, 
up to isometries.

In coordinates $(t,r,\theta, \varphi)$ adapted to the stationarity and axi-symmetry vector fields, $\partial/\partial t$ and $\partial/\partial \varphi$ respectively, this definition requires periodicity only in ($r,\theta$). Generically, LRs can be determined  via the $h_\pm(r,\theta)$ functions defined in~\cite{Cunha:2016bjh}. A LR is either a saddle point or an extremum of these functions, for fixed ($r,\theta$). The analogue of spherical orbits in non-separable spacetimes, however, is meaningless, since $r=const.$ is not preserved by mixing $r$ and $\theta$, and no key property, such as separability, singles out a particular coordinate chart. 

\begin{widetext}

\begin{figure}[t!]
\begin{center}
\includegraphics[width=0.265\textwidth]{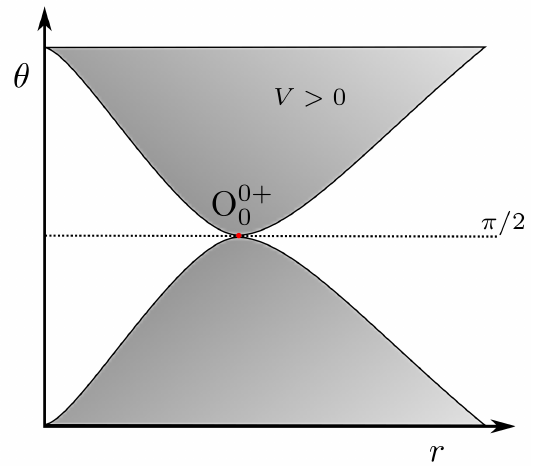}
\includegraphics[width=0.68\textwidth]{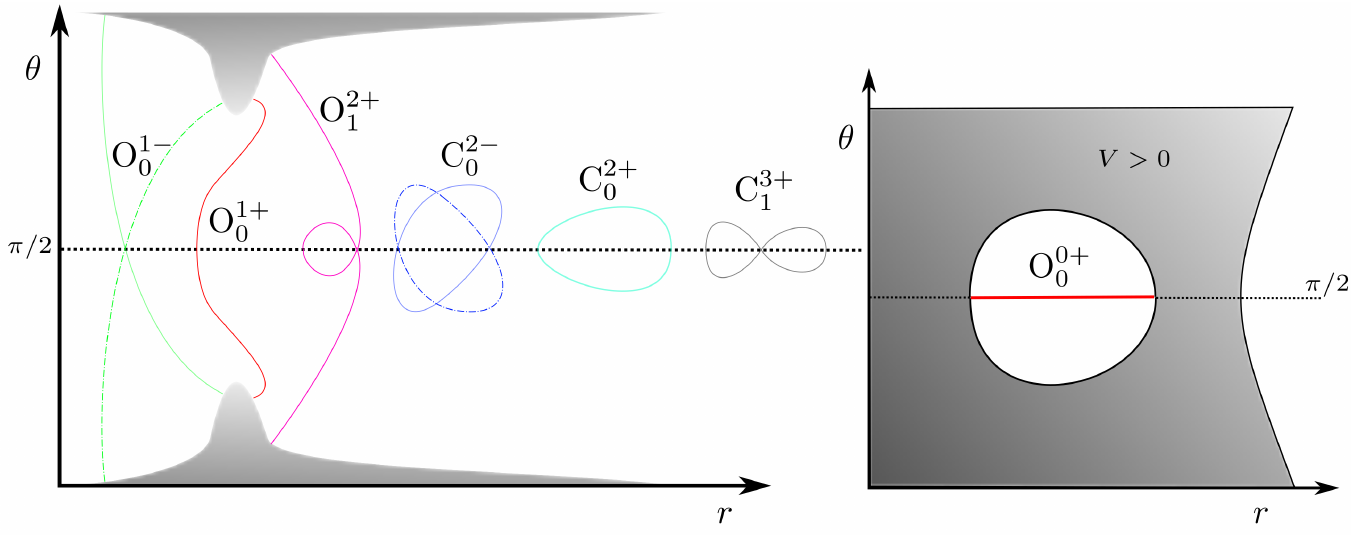}
\end{center}
\caption{\small Illustration of some  FPOs in the $(r,\theta)$-plane and their classification. The grey areas represent forbidden regions with $V>0$. The left/right panels show a typical unstable LR and a stable planar orbit.}
\label{class}
\end{figure}

\end{widetext}

\noindent{\bf {\em Classification.}}  The null geodesic flow on a spacetime $(\mathcal{M},g_{\mu\nu})$ is described by the Hamiltonian $\mathcal{H}=\frac{1}{2}g^{\mu\nu}p_\mu\,p_\nu=0$, where $p_\mu$ is the photon's 4-momentum. Besides stationarity, axi-symmetry and asymptotic flatness, with the metric expressed in the aforementioned coordinates,  we further assume a $\mathbb{Z}_2$ reflection symmetry on the equatorial plane $(\theta=\pi/2)$ and metric invariance under the simultaneous reflection $t\to-t$ and $\varphi\to-\varphi$~\footnote{Gauge freedom is used to set $g_{r\theta}=0$.}.

In terms of the first integrals $p_t\equiv -E$ and $\Phi\equiv p_\varphi$, we define a potential $V(r,\theta)$ and a kinetic term $T\geqslant 0$~\cite{Cunha:2016bjh}:
\[0=2\mathcal{H} = \underbrace{g^{rr}{p_r}^2 + g^{\theta\theta}{p_\theta}^2 }_{T\geqslant 0} + \underbrace{g^{tt}E^2 - 2g^{t\varphi}E\,\Phi + g^{\varphi\varphi}\Phi^2}_{V \leqslant 0} \ .\]
$V>0$ defines a forbidden region in phase space. At its boundary, $V=0 \Rightarrow  p_r=0= p_\theta$. From Hamilton's equations, $\dot{p}_\mu=-\frac{1}{2}\left(\partial_\mu g^{rr}p_r^2 + \partial_\mu g^{\theta\theta}p_\theta^2 + \partial_\mu V\right)$~\footnote{$\dot{p}_\mu$ denotes derivative with respect to an affine parameter.}.
The limit $V\to 0$ leads to $\dot{p}_\mu \to -\frac{1}{2}\,\partial_\mu V$. Hence, photons can only hit the boundary of the allowed region  $(V=0)$ perpendicularly. The null geodesic flow only depends on an impact parameter $\eta\equiv \Phi/E$; fixing $\eta$ determines the boundary of the forbidden region $V=0$. 

Within this setup, we categorized FPOs as $X^{n_r \pm}_{n_{s}}$, where $X=\{O,C\}$, and $n_r,n_{s}\in \mathbb{N}_0$: \\
{\bf i)} they either reach the boundary [\textit{class O} (open)], or they do not [\textit{class C} (closed)], in which case they loop;\\ 
{\bf ii)} they are either even (\textit{subclass$^+$}) or odd (\textit{subclass$^-$}) under the  $\mathbb{Z}_2$ reflection symmetry. For odd states a distinct mirror orbit exists; \\
{\bf iii)} they cross the equatorial plane ($\theta=\pi/2$) at $n_r$ distinct $r$ values (subclass$^{n_r}$). Orbits \textit{on} the equatorial plane, such as LRs, have $n_r=0$ (they never cross it);\\
{\bf iv)} They have $n_{s}$ self-intersection points  (subclass$_{n_{s}}$).\\

Some illustrations of these orbits are given in Fig.~\ref{class}. Typical LRs and more generic planar orbits are type $O^{0+}_0$ (left and right panels). Examples of the latter have been found, $e.g.$ in~\cite{Cunha:2016bjh}. $\mathbb{Z}_2$ odd orbits, such as $O^{0-}_0$, exist for instance in the $\mathbb{Z}_2$ Majumdar-Papapetrou dihole~\cite{Shipley:2016omi}. The Kerr FPOs are all of class $O^{1+}_0$. We have verified class $O^{2+}_1$ and $C^{2+}_0$ exist for rotating Proca stars~\cite{Brito:2015pxa}.

\noindent{\bf {\em Stability.}} 
The stability of FPOs can be analysed with  \textit{Poincar\'e maps} (see $e.g.$~\cite{josesaletan}). The relevant phase space is the 4-dimensional manifold $\mathbb{M}$, parameterized by $(r,\theta, \dot{r},\dot{\theta})$. Consider a null geodesic $s$ on $\mathbb{M}$ and let $\mathbb{P}$ be a  \textit{Poincar\'e section}, a submanifold of $\mathbb{M}$, which is assumed to intersect $s$ at multiple points. Usually the dimension of $\mathbb{P}$ is taken to be $\dim(\mathbb{M})-1=3$, but since there is an additional Hamiltonian constrain, we consider  dim($\mathbb{P}$)=2. A \textit{Poincar\'e map} $f:\mathbb{P}\,\to\,\mathbb{P}$, sends a given point of intersection with $s$ to the next intersection point. Parameterising  $\mathbb{P}$ by ${\bf x}=\{x^1,x^2\}$, the Poincar\'e map reads $f({\bf x}_n)={\bf x}_{n+1}$. This defines a discrete sequence of the intersection points, indexed by $n$.

 For a FPO, it is always possible to find $\mathbb{P}$ having fixed points $\tilde{{\bf x}}$ of this map, at which $f(\tilde{\bf x})=\tilde{\bf x}$. Its stability is determined by the behaviour of $f$ in the neighbourhood of $\tilde{\bf x}$. Taylor expanding to first order reads $f({\bf x}_n)\simeq f(\tilde{\bf x}) + \nabla f(\tilde{\bf x})\cdot {\bf y}_n$, 
where  $\nabla f$ is a $2\times 2$ matrix $A_{kj}\equiv(\nabla f^k)_j = \partial_jf^k$ and 
 ${\bf y}_n\equiv {\bf x}_n - \tilde{\bf x}$ is the deviation variable. Neglecting the higher order terms, ${\bf y}_{n+1}\simeq \nabla f(\tilde{\bf x})\cdot {\bf y}_n$, such that the $N^{th}$ term of a sequence starting with a deviation ${\bf y}_0$ is $ {\bf y}_N \simeq \left[\nabla f(\tilde{\bf x})\right]^N\cdot {\bf y}_0$. The value of ${\bf y}_N$ may diverge depending on the properties of (the matrix) $\nabla f(\tilde{\bf x})$, and in particular, of the modulus of its eigenvalues $\Lambda_k$: if $|\Lambda_k|\leqslant 1, \textrm{for all }k$, the orbit is stable; if  $|\Lambda_k|> 1, \textrm{for at least one }k$, the orbit is unstable.

Consider O$^{1+}_{0}$ orbits and let $\mathbb{P}$  be the equator $\theta=\pi/2$. Using the Hamiltonian constraint, a local patch of $\mathbb{P}$ is parametrized by ${\bf x}=(r,\dot{r})$. At the fixed point, $\tilde{{\bf x}}=(\tilde{r},0)$, only two (symmetric) values of $\dot{\theta}$ are possible. For simplicity,  restrict $\mathbb{P}$ to include only the fixed point with $\dot{\theta}\geqslant 0$~\footnote{The O$^{1+}_{0}$ actually intersects $\mathbb{P}$ with a symmetric $\dot{\theta}$ before returning to the initial point on $\mathbb{P}$. However, we could then redefine the map $f(x)\to (f\circ f)(x)$, so that $f(\tilde{x})=\tilde{x}$ without an intermediate point.}.  Defining $D=\det(A)$ and $T=\textrm{trace}(A)/2$, the eigenvalues are $\Lambda_\pm=T \pm \sqrt{T^2-D}$. For Hamiltonian systems $D=\pm 1$~\cite{josesaletan}. The  examples below have $D=1$ and fall into one of two cases. If $T^2> 1$, one of the eigenvalues has modulus larger than unity, and the orbit is unstable. If $T^2\leqslant 1$, the eigenvalues $\Lambda_\pm=T\pm i \sqrt{1-T^2}$ have unit modulus, leading to a rotation of the Poincar\'e map around the fixed point, which is therefore stable~\footnote{This analysis provides a simple criterion for the stability of the FPO.  $D=-1$ was not found within the cases analysed herein. In such case, the orbit is unstable for $T\neq 0$, and stable for $T=0$}.

\noindent{\bf {\em Kerr (and Kerr-like) FPOs.}} A generic Kerr solution has two LRs (see $e.g.$~\cite{Bardeen:1972fi}), one for a negative impact parameter, $\eta_-^{LR}$, and the other for a positive one, $\eta_+^{LR}$~\footnote{$\eta$ is always presented in units of the ADM mass $M$.}. The specific value of $\eta_\pm^{LR}$ depends on the BH spin. A continuum of FPOs exists with $\eta_-^{LR}<\eta<\eta_+^{LR}$. Each of these is, in BL coordinates, a spherical orbit that crosses the equatorial plane at a given perimetral radius, $r_{\rm Peri}$~\footnote{$r_{\rm Peri}$  is  defined  such  that  2$\pi r_{\rm Peri}=\oint d\varphi \sqrt{g_{\varphi\varphi}}$, where the metric component $g_{\varphi\varphi}$ is taken at a spacelike slice and on the equatorial plane, and $\partial /\partial \varphi$ is
the azimuthal Killing vector field.}, in between those of the two LRs, and attains a maximal/minimal angular coordinate $\theta_{\rm max}$. Observe that $\theta_{\rm max}=0,\pi$ for $\eta=0$, such that $\Delta\theta\equiv |\theta_{\rm max}-\pi/2|$ reaches $\pi/2$. The FPO with $\eta=0$ is actually the only complete spherical orbit; the remaining ones fail to reach high latitudes - Fig.~\ref{kerrfig} (left and middle panels).

All Kerr FPOs are unstable ($T^2>зк 1$). Neighbouring orbits to FPOs either escape to infinity or fall into the BH. Hence, these unstable FPOs determine the edge of the BH shadow - Fig.~\ref{kerrfig} (right panel). Rotating BHs in modified gravity (or in GR with reasonable matter contents) have typically small deviations from Kerr, including in their shadows.  Thus a similar picture for FPOs holds for many  rotating BHs, leading, in particular, to (qualitatively) Kerr-like shadows. Examples exist both in GR and beyond GR~\cite{2000CQGra..17..123D,Amarilla:2010zq,Amarilla:2011fx,Amarilla:2013sj,Grenzebach:2014fha,Grenzebach:2015oea,Cunha:2016wzk,Abdujabbarov:2016hnw,Mureika:2016efo,Dastan:2016bfy}.

\noindent{\bf {\em Non-Kerr FPOs.}} Significant non-spherical deformation of the Schwarzschild BH can lead to exotic features in its optical images~\cite{Abdolrahimi:2015kma}. For rotating BHs arising in a reasonable GR model with energy conditions abiding matter, non-Kerr-like shadows have been reported~\cite{Cunha:2015yba} for Kerr BHs with scalar hair~\cite{Herdeiro:2014goa,Herdeiro:2015gia}. Here, we  illustrate non-Kerrness using a ``cousin" model: Kerr BHs with Proca hair~\cite{Herdeiro:2016tmi}. In these hairy BHs, the null geodesic flow is non-integrable and chaos occurs for some (sufficiently) hairy BHs~\cite{Cunha:2016bjh}. Recent work suggests the dynamical formation of Kerr BHs with Proca hair~\cite{East:2017ovw}, justifying a detailed analysis of the theoretical and phenomenological properties of this family of solutions. 

\begin{figure}[b!]
\begin{center}
\includegraphics[width=0.238\textwidth]{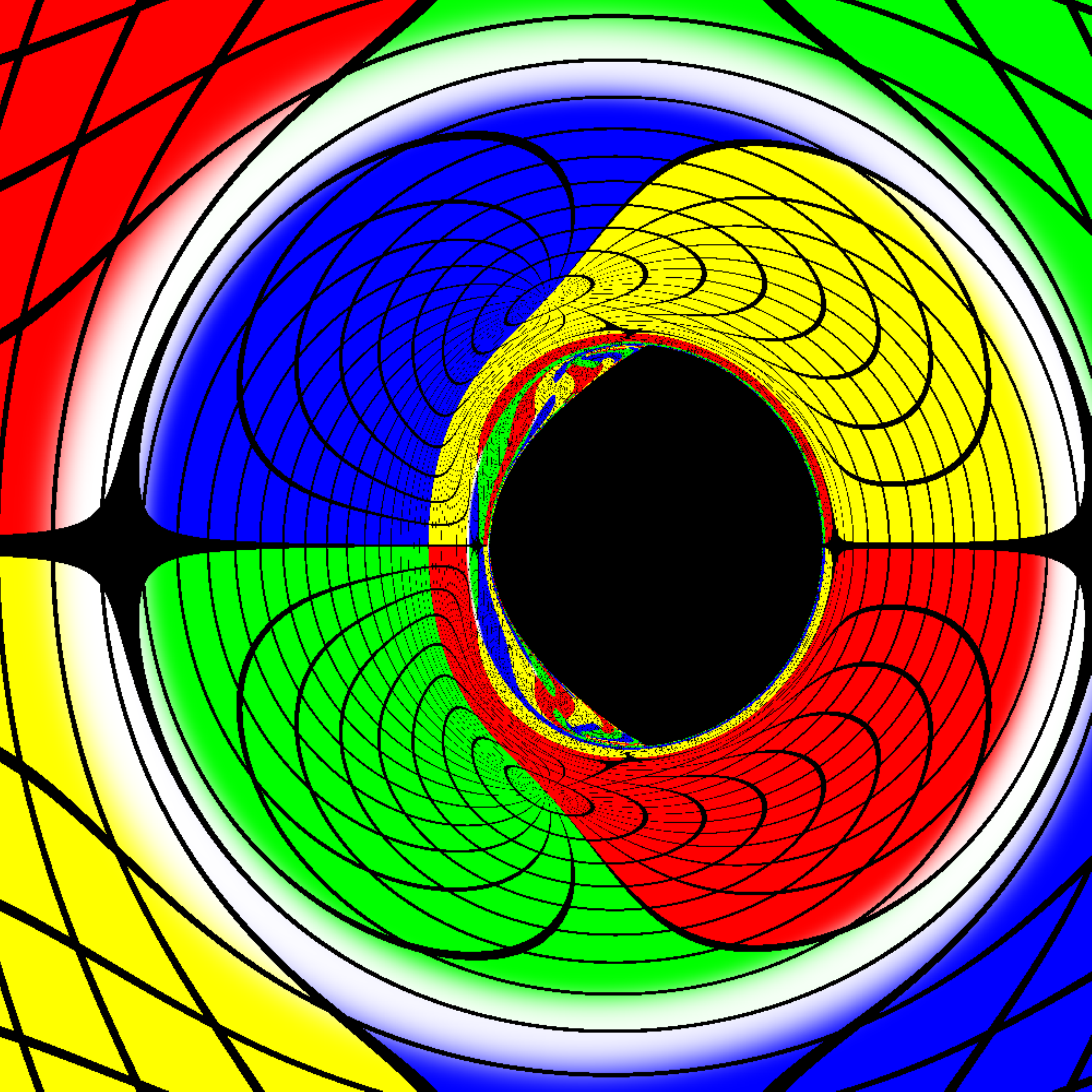}
\includegraphics[width=0.234\textwidth]{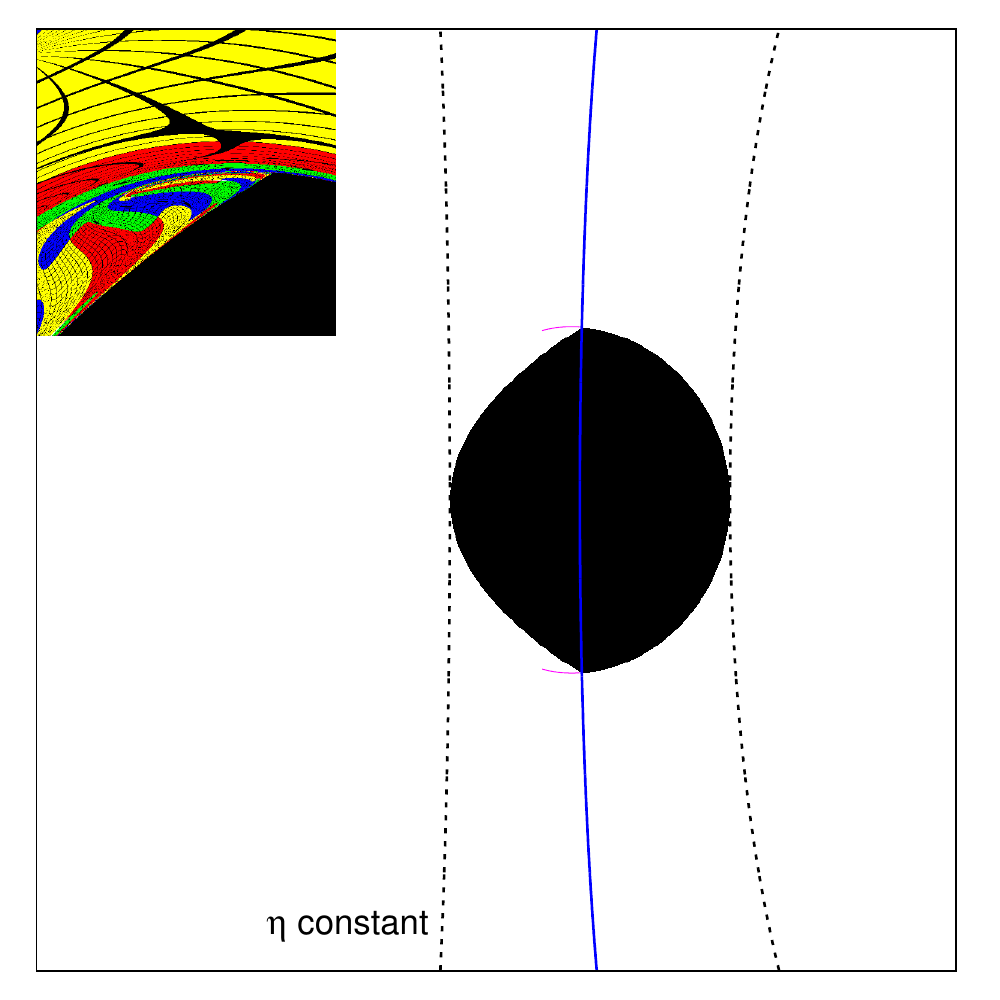}
\caption{(Left panel) Lensing of the hairy BH with a cuspy shadow, obtained with the same setup as in~\cite{Cunha:2015yba}. (Right panel) The cuspy shadow in the same observation conditions as the ones for the Kerr BH [which has the same $(M,J)$] in Fig.~\ref{kerrfig}. Almost vertical lines have constant $\eta$ and in this case there is no analogue of the Carter's constant. The small (pink) eye lashes correspond to a particular lensing pattern connecting to the cusp, which can be observed in the inset.}
\label{shadowsnk}
\end{center}
\end{figure}

Amongst these hairy BHs we have chosen a solution which is a sharp and illustrative example  of (non-Kerr-like) FPOs, including stable ones. Its lensing produces the \textit{cuspy shadow} -- Fig.~\ref{shadowsnk}~\footnote{The corresponding Kerr BH with Proca hair has ADM [horizon] mass and angular momentum $(M,J)=(1.075,0.948)$ [$(M_H,J_H)=(0.045,0.012)$]. The Proca field oscillates with frequency $w=0.8$. All these quantities are in units of the Proca mass.}. The solution's (ADM) quantities, $M,J$, match those of the Kerr BH shown in Fig.~\ref{kerrfig}. This is a (very) hairy BH with $\sim$ 96\% of the mass and $\sim$ 99\% of the spin stored in the ``hair" (Proca field).

\begin{widetext}

\begin{figure}[h!]
\begin{center}
\includegraphics[width=0.97\textwidth]{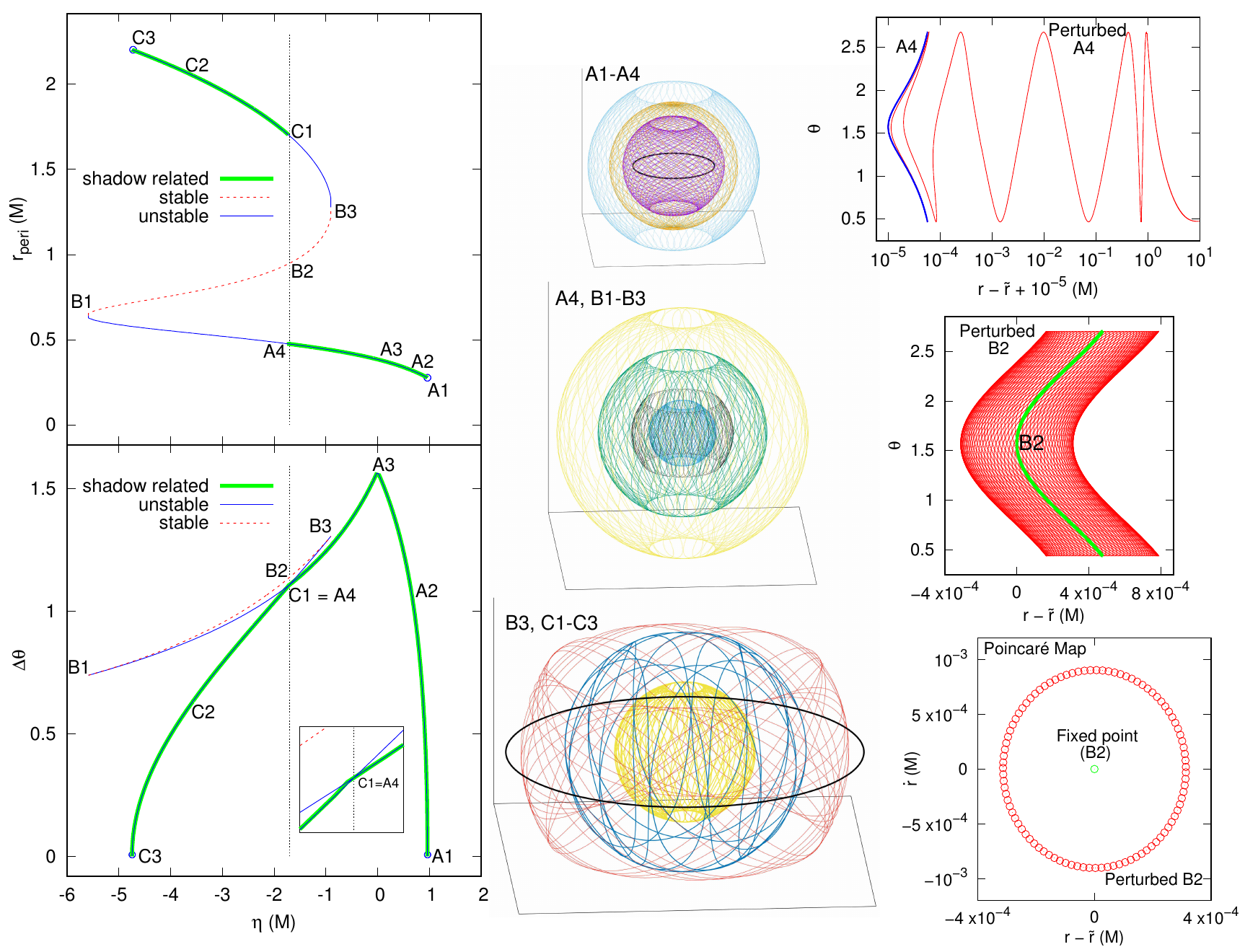}
\end{center}
\caption{\small Non-Kerr(-like) FPOs, illustrated for the hairy BH described in the text. (Left panel) $r_{\rm Peri}$ and $\Delta \theta$ for FPOs $vs.$ $\eta$.  We selected 10 FPOs (A1-A4,B1-B3,C1-C3), including the two LRs. The line  $\eta\simeq-1.71$ takes the value at which the cusp in the shadow occurs -- Fig.~\ref{shadowsnk}. (Middle panel) Spatial trajectories of these 10 FPOs, in Cartesian coordinates defined from the spheroidal coordinates in~\cite{Herdeiro:2016tmi}. The A4 (blue) and B3 orbits (yellow), at the intersection between stable and unstable branches are repeated to convey a sense of scale. (Right panel) One unstable [stable] FPO of the group A (top) [B (middle)] and a neighbouring perturbed orbit which diverges from [oscillates around] the FPO, together with the Poincar\'e map (on $\theta=\pi/2$) of B2, showing rotation about the fixed point $(r,\dot{r})=(\tilde{r},0)$.}
\label{nkfig}
\end{figure}

\end{widetext}

The salient feature of the cuspy shadow is its non-smooth edge. This feature, which occurs also for some Kerr BHs with scalar hair, is a consequence of the FPOs of this solution, as can be observed by analysing the $r_{\rm Peri}$ and $\Delta \theta$ for these FPOs, in terms of the impact parameter $\eta$ -- Fig.~\ref{nkfig} (left panel).

Fig.~\ref{nkfig} (left panel) informs us that, as for Kerr, there are two LRs, for $\eta_\pm^{LR}=-4.75;0.97$. However, differently from Kerr, these LRs are connected by a continuum of FPOs that can be split into three branches: two unstable (with $T^2>1$, that connect to the LRs) and a stable one, with $T^2\leqslant 1$, in between. A careful analysis of the two unstable branches reveals that only a part of each (green thicker lines) contributes to the edge of the shadow. The remaining unstable FPOs, as well as the stable FPOs, do not. Since the edge of the shadow on the equatorial plane is determined by the LRs, the FPOs that determine this edge must jump between the two branches. The jump occurs at the FPOs C1 and A4, which have the same  $\eta\simeq-1.71 M$ and attain the same angular deviation $\Delta \theta$. But there is a discontinuity in the size of these orbits, $r_{\rm Peri}(C1)>r_{\rm Peri}(A4)$, inducing the cusp in the shadow, precisely at $\eta\simeq-1.71$ (Fig.~\ref{shadowsnk}, right panel, blue line).

The unstable FPOs that are not associated to the shadow edge can, however, impact on the lensing properties of the spacetime. This is manifest in the \textit{eye lashes} depicted in Fig.~\ref{shadowsnk} (right panel, pink lines) which are associated to FPOs between C1 and B3, and form a clear lensing pattern (inset): a \textit{ghost shadow edge} from that branch of unstable FPOs. Finally,  if any photon bound orbit induces a spacetime non-linear  instability~\cite{Keir:2014oka,Cardoso:2014sna}, such instabilities would be missed by analysing solely LRs. Indeed, this example illustrates that non-planar stable FPOs may exist  \textit{without} planar ones (LRs). 

\noindent{\bf {\em Remarks.}} FPOs are the generic counterpart of LRs in a stationary, axisymmetric spacetime (see~\cite{Shipley:2016omi,Yoshino:2017gqv} for other discussions on extension of LRs). The illustrations herein show that FPOs can have a richer structure than in Kerr, and are instrumental in understanding BH shadows, lensing properties and spacetime stability. Thus, general FPOs can yield spacetime information beyond the scope of LRs. An extension of this concept, for generic spacetimes without any isometries, such as dynamical BH binaries, would be of interest.

\noindent{\bf {\em Acknowledgements.}}
P.C. is supported by Grant No. PD/BD/114071/2015 under the FCT-IDPASC Portugal Ph.D. program.
C. H. and E. R. acknowledge funding from the FCT-IF programme.  This work was partially supported by
the H2020-MSCA-RISE-2015 Grant No. StronGrHEP-690904,  and by the CIDMA project
UID/MAT/04106/2013. Computations were performed at the Blafis cluster, in Aveiro University.


\newpage

\bibliography{letter_shadows}

 
\end{document}